\documentclass[12pt]{iopart}
\newcommand{\be}{\begin{equation}}
\newcommand{\ee}{\end{equation}}
\newcommand{\bea}{\begin{eqnarray}}
\newcommand{\eea}{\end{eqnarray}}
\newcommand{\bt}{\begin{tabular}}
\newcommand{\et}{\end{tabular}}
\newcommand{\lessim}{\; ^< \!\!\!\! _\sim \;}
\newcommand{\neff}{N_{\rm eff}}
\begin{document}
\submitto{Physica Scripta \hfill MPP-2006-52}

\title{Standard and non-standard primordial neutrinos}

\author{P. D. Serpico}

\address{Max-Planck-Institut f\"{u}r Physik,
(Werner-Heisenberg-Institut),
\\F\"{o}hringer Ring 6, 80805, Munich, Germany}
\ead{serpico@mppmu.mpg.de}
\begin{abstract}
The standard cosmological model predicts the existence of a cosmic
neutrino background (C$\nu$B) with a present density of about
$n_\nu+n_{\bar{\nu}}\simeq 110\,$cm$^{-3}$ per flavour, which
affects big-bang nucleosynthesis (BBN), cosmic microwave background
(CMB) anisotropies, and the evolution of large scale structures
(LSS). We report on a precision calculation of the C$\nu$B
properties including the modification introduced by neutrino
oscillations. The role of a possible neutrino-antineutrino asymmetry
and the impact of non-standard neutrino-electron interactions on the
C$\nu$B are also briefly discussed.
\end{abstract}
\noindent\pacs{14.60.Lm, 98.80.-k \hfill {\it Keywords}: neutrinos,
cosmology}
\section{Neutrino decoupling including oscillations}
In the early universe, thermally produced neutrinos were in
equilibrium with other particles down to temperatures of few MeV,
when weak interactions became ineffective and neutrino decoupled
from the plasma (for general and updated reviews on the C$\nu$B and
its impact on cosmology, the reader is addressed
to~\cite{review1,review2}). As a first approximation, the
thermal Fermi-Dirac momentum spectrum is preserved after
the freeze-out of weak interactions, since neutrinos decoupled when
ultra-relativistic and both neutrino momenta and temperature
redshift identically with the universe expansion. Shortly after
neutrino decoupling, the photon temperature dropped below the
electron mass, and the $e^{\pm}$ annihilations heated the photons,
whose temperature was thus raised with respect to neutrino one by a
factor $z_{\rm fin}^0\equiv(11/4)^{1/3}\simeq 1.4010$ in the limit
of instantaneous neutrino decoupling. However, neutrino decoupling
and $e^{\pm}$ annihilations were sufficiently close in time that
some relic interactions between $e^{\pm}$ and neutrinos existed,
especially in the UV tail of the distributions. One expects then a
slightly smaller increase of the comoving photon temperature and
non-thermal distortions in the neutrino spectra, larger for $\nu_e$
and $\bar{\nu}_e$, which also undergo charged current interactions
with $e^{\pm}$. In~\cite{Mangano:2005cc}, we studied this process in
detail, by solving numerically the momentum-dependent density matrix
equations for the neutrino spectra, including finite temperature QED
corrections to the electromagnetic plasma, and for the first time in
an exact way the effect of three-neutrino flavour oscillations.

The global contribution of neutrinos to the energy density is
usually parameterized via an effective number of neutrinos $\neff$
(=3 for instantaneous decoupling),

\be \neff
\equiv\frac{\rho_{\nu+x}}{\rho_\nu^0}\frac{\rho_\gamma^0}{\rho_\gamma}
=\left(3+\sum_\alpha\delta_\alpha\right)\left(\frac{z_0}{z}\right)^4
\label{neff1},
 \ee
where $\rho_\gamma^0$ and $\rho_\nu^0$ are respectively the energy
density of the photon plasma and of a single neutrino species in the
limit of instantaneous decoupling, $\rho_\gamma$ the actual energy
content of the photon plasma, and $\rho_{\nu+x}$ the total energy
content of weakly interacting particles (including possible exotic
contributions). The second equality follows when only the three
active neutrinos contribute to $\rho_{\nu+x}$; there, the different
photon temperature evolution accounts for the second factor and the
relative energy-density distortion in the $\alpha$-th neutrino
flavour is given by
$\delta_\alpha\equiv(\rho_{\nu_\alpha}-\rho_\nu^0)/\rho_\nu^0$. In
Tab.~\ref{tab:mixed} we report our results for neutrino distortion
observables for several cases: the one without oscillations nor QED
corrections to the plasma, the one adding only QED corrections, and
three cases including oscillations, with the first one close to
present best fit of neutrino world data.
\begin{table}
\begin{center}
\begin{tabular}{lcccccc}
\hline Case & $z_{\rm fin}$ & $\delta_e$(\%) & $\delta_\mu$(\%) &
$\delta_\tau$(\%) & $\neff$ & $\Delta Y_p$\\
\hline No mixing, no QED & 1.3990 & 0.95 & 0.43& 0.43 & 3.035
&1.47${\times} 10^{-4}$\\
No mixing & 1.3978 & 0.94 & 0.43& 0.43 & 3.046&$ 1.71{\times} 10^{-4}$\\
$\{\theta_{12},\theta_{23},\theta_{13}\}=33^\circ,45^\circ,0^\circ$
& 1.3978 & 0.73 & 0.52 & 0.52 & 3.046 &$2.07 {\times} 10^{-4}$\\
$\{\theta_{12},\theta_{23},\theta_{13}\}=33^\circ,45^\circ,12.5^\circ$
& 1.3978 & 0.70 & 0.56
& 0.52 & 3.046 &$2.12 {\times} 10^{-4}$\\
$\{\theta_{12},\theta_{23},\theta_{13}\}=45^\circ,45^\circ,0^\circ$
&
1.3978 & 0.69 & 0.54 & 0.54 & 3.045 &$2.13 {\times} 10^{-4}$\\
\hline
\end{tabular}
\caption{Frozen values of the photon to neutrino ``temperature
ratio'' $z_{\rm fin}$, the energy density corrections
$\delta_\alpha$, the parameter $N_{\rm eff}$ and the change in the
$^4$He mass fraction $\Delta Y_p$ for different cases.}
\label{tab:mixed}
\end{center}
\end{table}
We find that, differently from QED effects, oscillations do not
essentially modify the total change in the neutrino energy density,
while the small effect on the production of primordial $^4$He is
increased by about 20\%: BBN is indeed sensitive separately to
$\delta_e$ and $\neff$, and removes the degeneracy. The results are
stable within the presently favoured region of neutrino mixing
parameters. Note that from the distortions in the C$\nu$B one would
expect ${\cal O}$(1\%) change in the predicted helium fraction
$Y_p$. However, the neutrino spectral distortions and the modified
photon temperature evolution conspire to almost cancel each other,
and $\Delta Y_p/Y_p\simeq  {\cal O}$(0.1\%) only. Nonetheless, this
effect is of the same order of the predicted uncertainty coming from
the error on the measured neutron lifetime and has to be included in
precise BBN predictions (see e.g.~\cite{Serpico:2004gx}).

\section{Neutrino asymmetry}
In the previous considerations we have neglected possible (comoving)
neutrino chemical potentials $\xi_\alpha$, that would exist in the
presence of a neutrino-antineutrino asymmetry. Is that a justified
approximation? Primordial nucleosynthesis yields are sensitive to
neutrino asymmetries  via: $a)$ the isospin-changing weak rates (a
positive $\xi_e$ enhances $n \rightarrow p$ processes with respect
to the inverse ones); $b)$ a modified expansion rate, since non-zero
$\xi_\alpha$'s contribute to total ${\rm N}_{\rm eff}$ as
\be \neff(\xi_\alpha)\simeq \neff(0) + \sum_\beta \left[
\frac{30}{7} \left(\frac{\xi_\beta}{\pi}\right)^2 + \frac{15}{7}
\left(\frac{\xi_\beta}{\pi}\right)^4 \right]. \label{eq:rhonudeg}
\ee
The former effect is dominant, in particular on $Y_p$, and a
stringent BBN bound on $\xi_e$ is obtained. Additionally, it was
shown in~\cite{Dolgov:2002ab} that neutrino oscillations lead to
flavour equilibrium before BBN: If neutrino chemical potentials are
non-vanishing, they should be equal to each other, and the BBN bound
on $\xi_e$ extends to all flavours. If no exotic degrees of freedom
are allowed in $\neff(0)$ and the CMB prior on the baryon abundance
is used, present constraints are at the level
$-0.05<\xi_e<0.07$~\cite{Serpico:2005bc}. It was however noted that
if both $\neff$(0) and $\xi_e$ are allowed to vary, the BBN bounds
on $\neff$ and $\xi_e$ are somewhat
relaxed~\cite{Barger:2003rt,Cuoco:2003cu}. Still $|\xi_e|\lessim
0.3$ holds, so that previous considerations and CMB and LSS
observables can not be significantly affected by a $\nu-\bar{\nu}$
asymmetry, at least within present accuracy of cosmological observations.

\section{Non-standard physics \& neutrino decoupling}
If flavour-changing or non-universal coupling to electrons are
present, they might delay the decoupling phenomenon: neutrinos get
more energy from $e^\pm$ annihilations. Although standard
non-thermal features are too small to be detected with present data,
larger non-thermal distortions could be more easily detected.
Phenomenologically, we can add to the standard model Lagrangian the
effective operators \be {\cal L}_{\rm
NSI}^{\alpha\beta}=-2\sqrt{2}G_F\sum_P
\epsilon_{\alpha\beta}^P(\bar{\nu}_{\alpha}\gamma^\mu L
\nu_\beta)(\bar{e}\gamma_\mu P e), \ee where
$P=L,R=(1\mp\gamma_5)/2$ and the dimensionless
$\epsilon_{\alpha\beta}^P$ parameterize the strength of non-standard
interactions (NSI). For example, preliminary calculations show that,
for $\epsilon_{\tau\tau}^L\simeq\epsilon_{\tau\tau}^R\simeq 2$,
$\neff$ as large as 3.17 and a variation in the $^4$He yield almost
of the order of 10$^{-3}$ can be obtained. When limiting the NSI to
vary within present laboratory bounds, one finds in general enhanced
non-thermal features, with a value of $(\neff-3)$ changed by up to a
factor $\sim 3$ with respect to the standard case. The effect of NSI
on the neutrino decoupling has been investigated in detail and the
results have been reported in \cite{Mangano:2006ar}.

\section{Conclusions}
Presently, C$\nu$B is starting being tested at a non-trivial level
by CMB and LSS data (see e.g.~\cite{Trotta:2004ty}), a progress
which is nicely reflected in the sharpening of the bounds on
$\neff$: although concerns still remain on possible unaccounted
systematics, it is intriguing that for the first time the
constraints quoted in \cite{Seljak:2006bg,Spergel:2006hy} are strong
enough---e.g. comparable with the traditional bounds from BBN---to
start discriminating among models. The PLANCK satellite temperature
and polarization data alone will probably pin-down this observable
to the level of $\Delta \neff =0.2$, and future missions might reach
the sensitivity of 0.04-0.05 needed to test the standard
scenario~\cite{Lopez99,Bowen:2001in,Bashinsky:2003tk}, possibly
ruling out many exotic models. Although no direct detection is
possible, we expect a bright future for the C$\nu$B!

\section*{Acknowledgments}
I thank the organizers of SNOW 2006 for the very pleasant and
stimulating workshop. It is a pleasure to thank G. Mangano, G.
Miele, S. Pastor, T. Pinto, O. Pisanti, and G. Raffelt for all
discussions and fruitful collaboration on the topics summarized in
this paper.

\end{document}